# Static Timing Model Extraction for Combinational Circuits

Bing Li, Christoph Knoth, Walter Schneider, Manuel Schmidt
Ulf Schlichtmann

Department of Electrical Engineering and Information Technology
Technische Universitaet Muenchen
Arcisstrasse 21, 80333 Munich, Germany
{b.li, christoph.knoth, manuel.schmidt, walter-karl.schneider,
ulf.schlichtmann}@tum.de

**Abstract.** For large circuits, static timing analysis (STA) needs to be performed in a hierarchical manner to achieve higher performance in arrival time propagation. In hierarchical STA, efficient and accurate timing models of sub-modules need to be created. We propose a timing model extraction method that significantly reduces the size of timing models without losing any accuracy by removing redundant timing information. Circuit components which do not contribute to the delay of any input to output pair are removed. The proposed method is deterministic. Compared to the original models, the numbers of edges and vertices of the resulting timing models are reduced by 84% and 85% on average, respectively, which are significantly more than the results achieved by other methods.

## 1 Introduction

Static timing analysis (STA) is widely used in digital circuit designs. The timing information of each circuit component is extracted from a design library. Combined with the result of parasitic extraction and the slopes of input signals, arrival times are propagated to compute the delays between the inputs and outputs of combinational circuits. Together with the timing characteristics of sequential components, e.g. the setup/hold time of flip-flops, the maximum/minimum of the computed delays determines the performance of the circuits.

In a hierarchical design flow, a design is composed of a series of circuit modules on different levels. To run static timing analysis for such a design style, the intuitive idea is to flatten all the sub-modules and run the propagation algorithm on this flat netlist. The disadvantage of this method is that the timing propagation through the full netlist is slow and very memory-consuming. Furthermore this flattening method will thwart the application of IP (Intellectual Property) macros from third-party vendors. The complete netlist of IP macros must be given to designers for static timing analysis, which is not always feasible for IP protection. In view of the reasons above, static timing analysis must also be hierarchical in such a design flow. As the first step, timing models of the sub-modules are generated from the original netlists. Then the arrival time propagation algorithm is run at circuit level, using the extracted timing models as the replacement of the netlists of the sub-modules. Naturally the extracted timing models must contain the exact delay information required by the high level analysis. To speed up the high level arrival time propagation, a timing model must be as compact as possible.

In [3] serial and parallel merges are introduced to reduce the size of combinational circuits and the compressed circuit netlists are used as timing models. The serial merge replaces two consecutive timing edges by a new one from the source of the first edge to the sink of the second edge. The parallel merge replaces parallel edges sharing the same source and sink by a single edge. Both operations are applied to the timing graph of the circuit iteratively until the size of the graph can not be reduced further. The application of these two fundamental operations heavily depends on the circuit structure and usually does not result in very small timing models for combinational circuits. In [5], a graph manipulation method is used to reduce the number of optimization constraints and timing variables in circuit optimization. In order to reduce the optimization complexity, circuit nodes whose removal can lead to fewer constraints and timing variables are deleted from the timing graph. This method transforms the circuit netlist targeting to effectively lower the optimization effort, but can not guarantee to generate a compact static timing model. In [2] a delay edge shrink method is used to merge several parallel delay edges followed by a single delay edge in series, and vice versa. Additionally, a parallel to serial graph transformation algorithm is also introduced to increase the possibility that the circuit can be compressed using the shrink method. These algorithms are applied to the original timing graph and the timing graph formed by directly connecting all the outputs of the original circuit with their driving inputs. Then, the smaller extracted timing model is selected as the final result. In [6] a timing model extraction method based on biclique-star replacement allowing don't care edges is introduced. This method extends the parallel to serial algorithm in [2] to deal with more than two inputs and outputs in the transformation. In order to find the proper replacement candidates for the biclique-star algorithm, an algorithm which maximizes edge coverage and the size of bicliques is also suggested in [6]. Both methods proposed in [2] and [6] are not deterministic and rely on heuristic pattern matching algorithms.

In this paper, we propose a method to extract timing models for combinational circuits by removing redundant timing information in the original timing graph. Without relying on heuristic pattern matching algorithms, the proposed method is deterministic and will always give compact results. Additionally, the proposed method does not use the graph manipulation algorithm in [2] and [6] and hence can be used as a pre-processing step before applying those algorithms.

In Section 2 we will formulate the problem and task of hierarchical timing model extraction. Then the proposed algorithms will be introduced in Section 3. In Section 4 we will show the extraction results of the proposed method using ISCAS85 benchmark circuits. Finally we will conclude our work.

## 2 Problem Formulation

In this section we will introduce the definition of a timing graph briefly. This timing graph will be used to illustrate our algorithms. Then, the requirement of the hierarchical timing model extraction for combinational circuits will be discussed. In the following sections, we will only discuss the worst-case timing model. The method for extracting the best-case timing model can be deduced similarly.

### 2.1 Timing Graph

A *timing graph* $G$ is a weighted directed graph representing the timing information of a circuit. A *vertex* $v_i$ corresponds to a net in the circuit. An *edge* $e_{ij}$ represents a delay between vertices $v_i$ and $v_j$. Each $e_{ij}$ has a *weight* $d_{ij}$, which is the magnitude of the delay between vertices $v_i$ and $v_j$. Compared with the circuit netlist, $e_{ij}$ corresponds to a pin to pin delay of a circuit component. When interconnects are considered, the delay of an interconnect is also represented by an edge. A *timing path* denoted as $p_{ij}$ is a set of consecutive connected edges between vertices $v_i$ and $v_j$. The path delay is the sum of the weights of all the edges on path $p_{ij}$ and is also denoted as $d_{ij}$. The set of all paths and their delays between vertices $v_i$ and $v_j$ are denoted as $P_{ij}$ and $D_{ij}$ respectively. A *sub-graph* is a set of edges from the timing graph and their source/sink vertices, denoted as $G\{e_{i_1j_1}, e_{i_2j_2}, \ldots\}$. In Fig. 1 the circuit c17 from the ISCAS85 benchmarks and its corresponding timing graph are illustrated without considering interconnects.

### 2.2 Formulation of the Hierarchical Timing Model

The target of static timing analysis is to compute the maximum delay between the inputs and the outputs of a circuit. In order to do this, a variable $a_i$ is assigned to each vertex $v_i$ in the timing graph, called *arrival time*. Then the timing graph is traversed by propagating the arrival times through all vertices until the output vertices are reached.

In hierarchical timing analysis, sub-modules are replaced by pre-characterized timing models. A *timing model* is a timing graph including a new set of internal edges and vertices but with the same inputs and outputs as the original timing graph. A timing model must contain the exact timing information needed by the higher level. Additionally, the timing model of a sub-module should be as small as possible to accelerate the arrival time propagation at high level. When characterizing a sub-module, especially an IP block, the application environment is unknown. For this reason, no assumption about the arrival times at the inputs of the sub-module should be made.

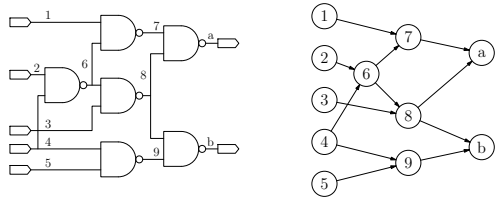 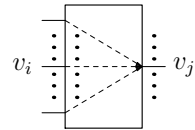

**Fig. 1.** c17 and its Timing Graph  **Fig. 2.** Arrival Time Computation of a Single Module

Fig. 2 illustrates the computation of the arrival time for an output $v_j$ of a sub-module during high level arrival time propagation. Firstly we consider the arrival time computation from only one input $v_i$ to the output $v_j$. Normally there is more than one path between $v_i$ and $v_j$, denoted as $P_{ij}$ with delay $D_{ij}$. The arrival time from $v_i$ through a path $p_{ij}$ to $v_j$ can be computed as $a_i+d_{ij}$. If we enumerate all the paths between $v_i$ and $v_j$, the maximum of these arrival times can be computed using (1), where $M_{ij}$ denotes the maximum path delay in $D_{ij}$. Now considering all the inputs of the module, the arrival time $a_j$ is the maximum arrival time from all the inputs to $v_j$. We get formula (2) to compute the arrival time $a_j$, where $V^I$ is the set of all the inputs of the module.

$$Max\{a_i + d_{ij}\} = a_i + Max\{d_{ij}\} = a_i + M_{ij}, \quad d_{ij} \in D_{ij} \qquad (1)$$

$$a_j = Max\{a_i + M_{ij}\}, \quad v_i \in V^I \qquad (2)$$

From (2) we can conclude that the arrival time at an output of a sub-module is determined by the arrival times at all the inputs and the maximum delays between the input/output pairs. The arrival times at the inputs of the module are computed during high level arrival time propagation and are still unknown when characterizing timing models. On the contrary, the maximum of the input to output delays $M_{ij}$ in (2) are provided by the timing model. In order to propagate arrival times correctly, a timing model must have the same input to output delays $M_{ij}$ as the original timing graph.

**Definition:** The *delay matrix* of a circuit with $m$ inputs and $n$ outputs is an $m \times n$ matrix with item $M_{ij}$, which is the maximum path delay between input $v_i$ and output $v_j$.

**Theorem:** The requirement for timing model extraction is that the characterized timing model must have the same delay matrix as the original timing graph.

## 3  The Timing Graph Reduction Algorithm

In this section we will explain a timing graph reduction method to extract compact timing models for combinational circuits. The basic edge shrink operation from [2] is introduced first. Thereafter two algorithms are explained which pre-process the original timing graph so that the basic shrink operation can be applied more effectively. Furthermore an output edge reduction algorithm will also be explained to compress the timing model. The target of the proposed timing model extraction method is to reduce the number of edges and vertices in the timing graph while maintaining the same delay matrix.

### 3.1  Basic Edge Shrink Operation

The basic edge shrink operation is illustrated in Fig. 3. If $n$ edges with sink vertices $v_{j_1} \ldots v_{j_n}$ leave the same vertex $v_k$ and $v_k$ has only one fanin edge with source vertex $v_i$, $v_k$ can be removed and the edges can be merged so that there are only direct edges between $v_i$ and $v_{j_1} \ldots v_{j_n}$. The weights of the new edges between $v_i$ and $v_{j_1} \ldots v_{j_n}$ are the sums of the weights $d_{ik}$ and $d_{kj_1} \ldots d_{kj_n}$, respectively. This shrink transformation is illustrated in Fig. 3(a). Similarly, this transformation can be applied in reverse direction, as shown in Fig. 3(b).

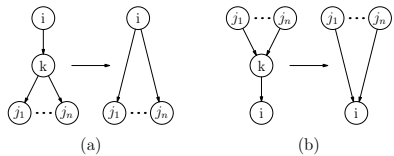
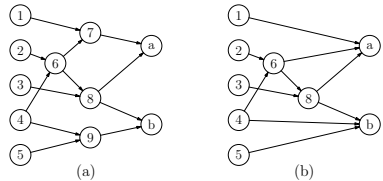

**Fig. 3.** Basic Shrink Operation

**Fig. 4.** Basic Shrink Example

Fig. 4 shows an example of applying the basic shrink operation to the timing graph in Fig. 1. Two sub-graphs G$\{e_{17}, e_{67}, e_{7a}\}$ and G$\{e_{49}, e_{59}, e_{9b}\}$ can be compressed using the basic shrink operation. After this reduction, the number of the edges of the timing graph is reduced from 12 to 10 and the number of the vertices is reduced from 11 to 9.

In Fig. 4(b), the timing graph can not be reduced further because no vertex except the inputs has only one fanin/fanout edge. In the following, we will propose several algorithms to pre-process the original timing graph so that the basic shrink operation can be applied more effectively to reduce the number of the edges and vertices.

### 3.2 Primary Input/Output Transformation (PIT/POT)

The principle of the timing graph reduction is that the delay matrix remains unchanged. A row in the delay matrix contains the maximal delays from one input to all the outputs, and a column contains the maximal delays from all the inputs to one output.

In many circuits with abundant datapaths there is delay symmetry, which means the rows or columns in the delay matrix share some patterns. If the subtraction of two rows or columns in the delay matrix is a vector whose elements are the same, the timing model needs to contain only the timing information for one of the rows or columns. This means both inputs or outputs can share the same delay paths in the timing graph and the edges and vertices constructing the delay paths for one of them can be deleted to simplify the timing graph.

If two rows $R_i$ and $R_j$ in the delay matrix have constant difference, that is, $R_i - R_j = C_r$, where $C_r$ is a row vector with constant element $c_r$, the two inputs $v_i$ and $v_j$ corresponding to $R_i$ and $R_j$ can be transformed by the algorithm illustrated in Fig. 5.

Assume the vertices $v_1$ and $v_2$ in Fig. 5 meet the condition $R_1 - R_2 = C_r$, we create a new vertex $v_{1'}$ between $v_1$ and all its fanout vertices. $v_{1'}$ is connected to all the fanout vertices of $v_1$ with the original edges from $v_1$. The new edge $e_{11'}$ has weight 0 so that the delays between $v_1$ and all the outputs do not change. In order to make $v_1$ and $v_2$ share the same timing paths, $v_2$ is disconnected from all its original fanout edges and a new edge between $v_2$ and $v_{1'}$ is created, as in Fig. 5(b). The weight $d_{21'}$ is set to $-c_r$ so that the delays from $v_2$ to all the outputs are also maintained because of $R_1 - R_2 = C_r$. After this transformation, the unconnected vertices and edges are deleted recursively, illustrated in Fig. 5(c). Thereafter there may be sub-graphs which can be reduced using the basic shrink operation, e.g. G$\{e_{35}, e_{56}, e_{57}\}$. These sub-graphs are compressed further, as shown in Fig. 5(d). By creating a new vertex $v_{1'}$ for the input transformation the necessity to create edges between $v_2$ and all the original fanout vertices of $v_1$ is avoided.

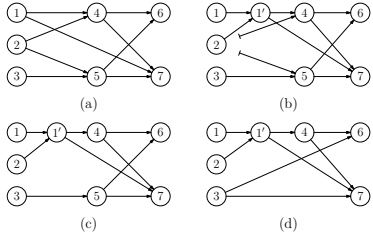
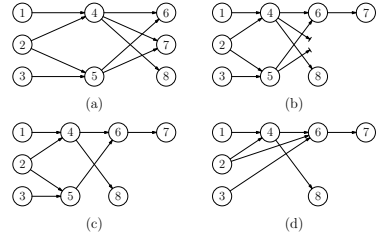

**Fig. 5.** Primary Input Transformation

**Fig. 6.** Primary Output Transformation

Similar to PIT, POT transforms the primary outputs of the original timing graph. If two columns $C_i$ and $C_j$ in the delay matrix meet $C_i - C_j = C_c$, where $C_c$ is a constant column vector with the element $c_c$, the output $v_j$ is disconnected from its driving edges and connected with the vertex of output $v_i$, through a new edge with weight $-c_c$. Similarly, the edges and vertices which originally drive only $v_j$ are deleted to reduce the graph size. The POT algorithm is illustrated in Fig. 6, where $v_7$ changes its connection to $v_6$. Because $v_6$ is an output in Fig. 6, it should not be deleted by applying the basic shrink operation to the sub-graph G$\{e_{26}, e_{36}, e_{46}, e_{67}\}$.

### 3.3 Non-Critical Edge/Vertex Removal (NCR)

Normally there is more than one path for an input/output pair in the original timing graph. In the view of the static timing analysis, only the path with the maximum delay, called *critical path* of the input/output pair, is needed in the high level arrival time propagation, as shown in (2). The other paths are dominated by the critical one and have no effect on the result of the timing analysis. From this observation, the edges which do not contribute to the delay matrix can be deleted to reduce the timing graph. Be aware that the definition of critical path here is different from the classical one, where the critical path dominates the paths starting from all the inputs of the timing graph. In our definition, the critical path dominates all the paths starting from a specified input to a specified output.

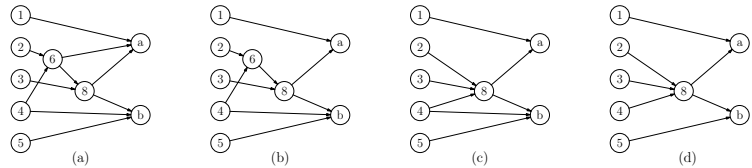

**Fig. 7.** Non-critical Path Removal

Fig. 7 illustrates the concept of the non-critical edge/vertex removal using the timing graph in Fig. 4(b). All the edge delays in Fig. 7 are assumed as unit for simplicity. Firstly, the redundant edge $e_{6a}$ directly between $v_6$ and $v_a$ is removed because there is another path through $v_6$, $v_8$ and $v_a$ with dominant delay, as shown in Fig. 7(b). After this removal a basic shrink operation can be applied to sub-graph $G\{e_{26}, e_{46}, e_{68}\}$ and the resulted timing graph is shown in Fig. 7(c). Similarly the redundant edge $e_{4b}$ can also be deleted. The final timing model is shown in Fig. 7(d).

To reduce the size of the timing graph, only the edges and vertices which are not on the critical path of any input/output pair can be removed safely. Because timing edges and vertices may be shared by the critical paths between different input/output pairs, we firstly traverse the timing graph and mark the vertices which are on at least one critical path. After identifying the critical paths for all the input/output pairs, the vertices and the edges which are never critical are deleted from the timing graph.

Instead of visiting all the input-output pairs one by one, whose number is the product of the number of the inputs $m$ and the number of the outputs $n$ in the worst case, the original timing graph is traversed only $m$ times [4]. At each traversal, the arrival times from a specified input to all the vertices are

```
Compute the arrival time from input v_i to all the outputs
```
```
levelize the circuit, maxlevel={the maximum circuit level}, currlevel=0
set the arrival time a_i of the input vertex v_i to 0
for each vertex in the timing graph except v_i
    set arrival time to -∞
add v_i in the vertex list of level 0
while (currlevel ≤ maxlevel) {
    for each vertex v_j in the vertex list of level currlevel {
        for all the fanin vertex v_k of v_j
            set a_j = Max{a_k + d_kj}                    (3)
        for all the fanout vertex v_s of v_j
            add v_s to the vertex list of the level of v_s
    }
    currlevel++
}
```
```
Mark the critical vertices from all outputs to the input v_i
```
```
for each output v_o {
    v_c = v_o, mark v_o as critical
    while (v_c is not an input and a_c > -∞) {
        v_c = {the fanin vertex v_k having the maximum a_k + d_kc}
        mark v_c as critical
    }
}
```

**Fig. 8.** Mark Critical Vertices for Input $v_i$ – singleMark($v_i$)

computed. We use the computed arrival times to trace the critical paths and mark the vertices on the critical paths backward. Fig. 8 shows the single input traversal and backward vertex mark algorithm-*singleMark()*.

Because the arrival times at the inputs except $v_i$ are set to $-\infty$ in Fig. 8, the arrival time of a vertex $v_j$ computed by (3) is the maximum delay from $v_i$ to $v_j$. To mark the critical vertices from the input $v_i$ to an output $v_o$, the algorithm traverses backward from $v_o$ to $v_i$ recursively. At each intermediate vertex $v_c$, the fanin vertex $v_k$ is marked as critical, when the sum of the arrival time $a_k$ and the delay $d_{kc}$ is the largest one in all the fanin vertices of $v_c$. $v_c$ is updated to $v_k$ for further backward traversal. At each run the tracing from an output to the input marks the path whose delay dominates the delays between the input and the output. All the edges and vertices on such a path will be kept in order to guarantee the conformability of the delay matrix.

The algorithm in Fig. 8 is applied to each input of the timing graph so that the vertices on the critical paths of all input/output pairs are marked. All the edges without marked fanin and fanout vertices do not contribute to the delay matrix. From the view of a timing model, these edges and unmarked vertices can be deleted to simplify the timing graph. Fig. 9 shows the complete non-critical edge/vertex removal algorithm.

```
for each input vertex v_i
   run singleMark(v_i)
for each vertex v_i in the timing graph
   if v_i is not marked
      delete v_i
for each edge e_i in the timing graph
   if e_i has no fanin or fanout vertex
      delete e_i
```

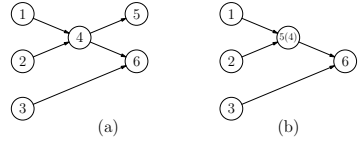

**Fig. 9.** Complete Algorithm of the Non-Critical Edge/Vertex Removal

**Fig. 10.** Output Backward Merge

### 3.4 Output Backward Merge (OBM)

If a primary output of a timing graph has only one fanin edge and its fanin vertex is not an output, we can simply merge this edge backwards. Fig. 10 illustrates this operation, where all the vertices except $v_4$ are primary inputs/outputs. In the OBM algorithm, vertex $v_5$ is merged with its fanin vertex $v_4$ so that the new vertex $v_{5(4)}$ represents also a primary output. All the weights of the fanin edges of $v_4$ are increased by $d_{45}$. This weight increase guarantees that the delays from $v_1$ and $v_2$ to the output $v_5$ remain unchanged. The weights of the other fanout edges of $v_4$ are all decreased with $d_{45}$ so that the delays from the inputs to the outputs besides $v_5$ are also unchanged. Note that the graph in Fig. 10(b) can not be reduced further using the basic shrink operation because $v_{5(4)}$ is already a primary output now and should be kept in the timing model. An application example of the OBM algorithm is the reduction of the sub-graph G$\{e_{14}, e_{24}, e_{46}, e_{48}\}$ in Fig. 6(d).

### 3.5 Complete Algorithm

```
apply PIT/POT to the timing graph
apply NCR algorithm to delete redundant vertices and edges
visit each vertex v_i starting from the inputs sequentially
   for sub-graph with v_i and its fanin/out edges and vertices
      run the basic shrink operation to the sub-graph
apply the output backward merge algorithm
```

**Fig. 11.** Complete Timing Model Extraction Method

## 4 Experimental Results

In this section the results of the application of the proposed method using the ISCAS85 benchmarks are shown. The algorithms are implemented in C++ and tested using a Pentium M 1.6GHz computer. The gates in the benchmarks are mapped to a library from an industrial partner. To model the impact of layout capacitance, the pin-to-pin delays of a gate are increased by 20% for each fanout. Interconnect delays are ignored for simplification.

Table 1 shows the extraction results. $m$ and $n$ are the numbers of the inputs and outputs of the circuit. $n_{e_o}$ and $n_{v_o}$ are the numbers of the edges and vertices in the original timing graph $G_o$. $n_{e_r}$ and $n_{v_r}$ denote the numbers of the edges and vertices of the final timing graph $G_r$ after applying the complete reduction method proposed in this paper. $p_{e_r}$ and $p_{v_r}$ are defined as $n_{e_r}/n_{e_o}$ and $n_{v_r}/n_{v_o}$ to show the ratios of the edges and vertices before and after the application of the proposed method. $T$ is the runtime of the proposed method. From Table 1 we can see that the proposed method compresses the timing graphs effectively.

According to [1] the benchmark c6288 is a 16×16 multiplier so that many critical paths are shared by different input to output pairs. This can explain why c6288 has the very drastic compression ratio. c499 and c1355 have the same circuit function, but all the XOR gates in c499 are expanded to NAND gates in c1355. This expansion increases the symmetry in the delay paths because smaller gates have better pin-to-pin delay symmetry than larger gates. This can explain why c1355 has a smaller timing model than c499.

| Circuit | Original circuit | | | | Proposed method | | | | | Results of [3] | | | | Results in [2] | |
|---|---|---|---|---|---|---|---|---|---|---|---|---|---|---|---|
| | $m$ | $n$ | $n_{e_o}$ | $n_{v_o}$ | $n_{e_r}$ | $n_{v_r}$ | $p_{e_r}$ | $p_{v_r}$ | $T$(s) | $n_{e_s}$ | $n_{v_s}$ | $p_{e_s}$ | $p_{v_s}$ | $n_{e_t}$ | $p_{e_t}$ |
| c432 | 36 | 7 | 336 | 196 | 41 | 42 | 12% | 21% | 0.06 | 211 | 82 | 63% | 42% | 65 | 19% |
| c499 | 41 | 32 | 408 | 243 | 138 | 67 | 34% | 28% | 0.12 | 240 | 99 | 59% | 41% | 175 | 43% |
| c880 | 60 | 26 | 729 | 443 | 212 | 103 | 29% | 23% | 0.17 | 331 | 126 | 45% | 28% | 238 | 33% |
| c1355 | 41 | 32 | 1064 | 587 | 106 | 67 | 10% | 11% | 0.25 | 240 | 99 | 23% | 17% | 147 | 14% |
| c1908 | 33 | 25 | 1498 | 913 | 165 | 76 | 11% | 8% | 0.25 | 456 | 126 | 30% | 14% | 337 | 22% |
| c2670 | 233 | 140 | 2076 | 1426 | 336 | 299 | 16% | 21% | 0.93 | 423 | 338 | 20% | 24% | 562 | 27% |
| c3540 | 50 | 22 | 2939 | 1719 | 327 | 109 | 11% | 6% | 0.69 | 1093 | 287 | 37% | 17% | 372 | 13% |
| c5315 | 178 | 123 | 4386 | 2485 | 884 | 377 | 20% | 15% | 1.56 | 1149 | 458 | 26% | 18% | 1109 | 25% |
| c6288 | 32 | 32 | 4800 | 2448 | 196 | 63 | 4% | 3% | 0.95 | 3313 | 1457 | 69% | 60% | 195 | 4% |
| c7552 | 207 | 108 | 6144 | 3719 | 946 | 510 | 15% | 14% | 2.78 | 1592 | 645 | 26% | 17% | 1717 | 28% |
| average | | | | | | | 16% | 15% | | | | 40% | 28% | | 23% |

**Table 1.** Comparison of the Reduction Results

For comparison we have implemented the serial/parallel method in [3]. Table. 1 shows the reduction results, where $n_{e_s}$ and $n_{v_s}$ are the numbers of the edges and vertices in the resulted timing graph respectively. $p_{e_s}$ and $p_{v_s}$ are defined as $n_{e_s}/n_{e_o}$ and $n_{v_s}/n_{v_o}$ respectively. In [2] only the edge numbers of the reduction results are given, which are listed as $n_{e_t}$ in Table. 1. $p_t$ is defined as $n_{e_t}/n_{e_o}$. Compared to our method, the method in [6] achieves slightly better edge compression ratios, but their resulting models involve error bounds and usually deviate from the original timing models. From these comparisons the efficiencies of the generated timing models by the proposed method are confirmed.

## 5   Conclusion

In this paper we proposed a method to effectively extract timing models for combinational circuits. After applying the primary input/output transformation and the non-critical edge/vertex removal algorithms, timing graphs are drastically compressed using the basic shrink operation recursively. Thereafter the primary output backward merge algorithm is applied to compress the timing graphs further. Compared to the original timing graphs, the numbers of edges and vertices of the resulting models are reduced by 84% and 85% on average, respectively. The extracted timing models have exactly the same delay matrix as their original timing graphs so that the accuracy of the high level arrival time propagation is maintained. The proposed method mainly focuses on the structural transformation of the original timing graph. Therefore, it can also be used as a pre-processing step before applying the method introduced in [2] and [6]. Future work will incorporate slope propagation and load effect into the timing model extraction.